**AntiPlag – A Sampling-based Tool for Plagiarism Detection in Student Texts**

Tuomo Kakkonen, Niko Myller
University of Joensuu, Finland
tuomo.kakkonen@cs.joensuu.fi
niko.myller@cs.joensuu.fi

**Abstract:** This paper introduces *AntiPlag*, an advanced plagiarism detection tool intended for use on student texts. It is capable of both *hermetic detection* that scrutinizes only local collections of documents (other students' texts and lecture materials, for example) and *web plagiarism detection,* in which the aim is at identifying instances of plagiarism that have been sourced from the Internet. The main feature of the system is the *sampling-based web plagiarism detection*, a novel approach to plagiarism detection that is based on combining web and hermetic search technologies. The system uses standard web search engines to locate documents on the Internet that might have been used as sources of plagiarism by the writer of a text. During this sampling phase, the suspected sources are downloaded, converted to ASCII text and saved to the local database so that they can be later processed by using the hermetic detection methods. We evaluated the system by using a test set that contained instances of verbatim copying as well as texts in which plagiarism was concealed by minor editing, replacing words with synonyms and by paraphrasing. We compared the results achieved by AntiPlag to an earlier evaluation study of four web plagiarism detection systems, SafeAssignment, TurnitIn, EVE2 and Plagiarism-Finder. AntiPlag performed better than any of these systems, achieving the accuracy 95.8% over all the test items.

**1. Introduction**

Many researchers and academics have noticed that the use of computers has resulted in an increase in student plagiarism. The use of ICT in teaching has caused an increase in the rate of plagiarism because of the wide availability of electronic texts. Many students are tempted to copy text from electronic sources such as course materials, the work of other students, and the Internet. In contrast to a *hermetic detection system* that scrutinizes only local collections of documents (other students' texts and lecture materials, for example), a *web plagiarism detection system* identifies instances of plagiarism from Internet sources.

Although automatic plagiarism detection systems are already available, experimental evaluations of these systems have shown that their accuracy in detecting plagiarism leaves a lot to be desired (Kakkonen & Mozgovoy 2008). Existing plagiarism detection systems, for example, have only a limited capacity to detection *web plagiarism*. This deficiency is attributable to one of the limitations of available web search services. They only return recognitions when one is searching for an exact phrase (such as "this is a query"). Currently available web search engines do not support wildcards (such as "this * a query"), or other, more flexile query techniques. Since most of the existing plagiarism detection systems do not use advanced search techniques or perform any significant natural language processing (NLP) (Uzuner et al (2005) and Mozgovoy et al. (2007) are among the few who have incorporated NLP techniques), they are often only capable of detecting direct copying and the kind of plagiarism that consists of simple editorial changes such as the addition, the removal or substitution of various characters. It has also been observed that the systems make false detections.

This paper introduces *AntiPlag*, an advanced plagiarism detection tool intended for use on student texts. The main research question in the development of AntiPlag and this paper has been:
- What is the most efficient way of detecting web plagiarism, i.e., how can Internet search engines and web documents effectively be utilized in AntiPlag?

AntiPlag tackles the problems that exist in current plagiarism detection systems in web plagiarism detection by combining web searchers with more advanced local search technologies. This is achieved by using web search engines for sampling purposes only. This method works by copying suspected sources of plagiarism into the local database where they are then subjected to the hermetic detection processes. We refer to this feature as *sampling-based web plagiarism detection.* While a combination of web and hermetic detection methods appears to be a feasible approach, the exact method in which these techniques are applied will be decided by research and empirical experiments. One of the main research problems is to determine an optimal method for incorporating web documents so that enough detection coverage and accuracy will be achieved without exploding the number of documents that need to be managed in the local database.

The following section (Section 2) introduces some of the existing plagiarism detection systems and their deficiencies. Section 3 outlines the architecture of the AntiPlag system. Section 4 provides details of the rigorous evaluation that we have carried out in order to test the detection accuracy and false detection rate of AntiPlag and to find out optimal parameters for sampling web pages for hermetic detection. The paper closes with Section 5, which concludes our findings.

**2. Background**

2.1 Plagiarism

During the past decade, student plagiarism has become a serious problem. The wide availability of the Internet and computers, and the ease, with which electronic texts can be copied, are among the most obvious factors that contribute to this problem. Posner (2007), for example, recently estimated that one third of high-school and college students have committed plagiarism at some time or another. This estimation might very well be an understatement. For example, 70.3% of the Finnish high-school and vocational school teachers that responded to our survey indicated that plagiarism is a problem in day-to-day teaching work (Kakkonen 2007). More than 80% of the teachers in the survey considered copying from the Internet to be the most commonly encountered form of plagiarism.

Park (2003) listed the following reasons as among the most common for committing acts of plagiarism: student ignorance about the correct way to acknowledge the use of referenced sources by means of proper citation and reference techniques, the desire to obtain better grades, the fact that many students are deficient in managing their time (poor time management skills), and, lastly, the perception among students that there is very little risk of being caught for having committed acts of plagiarism. Numerous styles of plagiarism exist. These range from the most common one, *verbatim copying*, to making use of a ghostwriter. Some of the most common forms of plagiarism are:
- "Copy-paste" verbatim copying
- Paraphrasing: the reordering of sentences and effecting changes to the words, grammar and style
- The insertion of deliberate errors: inserting spelling or grammatical mistakes into the source text
- The intentionally inaccurate use of references: the citing of incorrect or non-existent sources and the deliberate omission of sources

2.2 Fighting plagiarism

Posner (2007), among others, observes that plagiarism is a fascinating research subject because of the ambiguity inherent in the concept itself, the complexity of the relationships between plagiarism and other illicit copying practices (such as copyright infringement), differences of attitude towards plagiarism in different cultures, the complex motives and excuses of those who practice plagiarism, the diversity of the means of detecting plagiarism, and the kinds of punishment inflicted on plagiarists, and the forms of absolution that are made available to plagiarists in different contexts. This list makes it clear that countering student plagiarism is a complex issue that requires expertise in various branches of education, ethics and copyright law. It is essential, for example, to teach students how to reference their sources in accordance with widely accepted systems. It is also necessary to educate students to respect the honor codes that are being devised by educational institutions for the modeling of right conduct in matters of ethics. A regulation or law that forbids plagiarism will be useless if it is not able to be consistently and accurately enforced. The more sophisticated the methods for plagiarism detection are, the easier it will be to enforce ethical rules of academic conduct.

The motivation for fighting plagiarism is obvious. In the first instance, it prevents the educators to achieve their pedagogical goals. The mere copying of relevant texts is a low-grade skill that involves students in acts of moral compromise and deception and that is unfair to those students who are honest. The most common method of detecting plagiarism relies on the ability of an assessor to make deductions about the probability of plagiarism on the basis of internal clues embedded in the text itself. These clues include, for example, sudden changes in the vocabulary, style and grammar.

Countering plagiarism by using such "traditional" means is unfortunately unreliable and ineffectual. In addition, it wastes the valuable time of a teacher.

Although the ease of access to a great variety of texts on the Internet and other electronic sources has increased the incidence of plagiarism, it also offers educators various means with which to counter it. While Internet search engines, such as Google and Yahoo!, can be used to aid manual detection of web plagiarism, such a detection process is, by any standards, both tedious and labor-intensive.

2.3 Automatic plagiarism detection systems

Because the manual detection of plagiarism is so time-wasting and often inaccurate, automatic detection tools have been developed. A plagiarism detection system should ideally be able to detect all styles and instances of plagiarism while refraining from false detections, the stigmatization of portions of text in which no plagiarism has occurred. The value of using computerized tools (apart from the obvious value in detecting plagiarizers) is that they make students aware that there educational institutions are taking plagiarism seriously. This encourages students to learn and practice legitimate methods of citation and the referencing of their sources.

There are a number of companies, such as iParadigms (TurnItIn, http://www.iparadigms.com/) and Canexus (EVE2, http://www.canexus.com/), that have developed plagiarism detection software. While none of the companies that produce these detection systems provides information about the exact inner workings of their products, a great deal of published academic research (Wise 1996, Joy & Luck 1999, Schleimer et al. 2003) enables to make some inferences as to how the existing systems work. It is obvious, for example, that the most common methods used for hermetic detection involves the use of string matching algorithms. Web detection is typically based on the use of an existing web search engine such as Yahoo! or Google for searching the content of the Internet for parts of a document that might match a text turned in by a student.

Our recent evaluation results (Kakkonen & Mozgovoy 2008), however, show that existing plagiarism detection systems have serious deficiencies. While it has been noticed that there are gaps in the way that these systems scrutinize web pages and local documents, they are also prone to cite correctly referenced and cited sentences as plagiarisms. Most of these systems also find it difficult to detect cheating such as the use of false citations, or technical tricks such as the replacement of empty spaces with white-colored characters. Perhaps the most important limitation associated with existing systems is their limited capacity to detect forms of plagiarism in which the plagiarizer has effected changes to the source text. Yet another deficiency of these systems is that they may return false detections. Since the volume of texts on the Internet is so vast, it is possible that some part of a student's text will coincidentally resemble the text from an existing web page – even though the student might never have seen the web page concerned.

**3. AntiPlag**

AntiPlag is a plagiarism detection system that aims at high accuracy and low number of false detections. The initial version of AntiPlag was a Java application that used exact string matching and web search engines in order to perform hermetic and web plagiarism detection. It was capable of detecting verbatim copying from the Internet and from local documents (such as other students' essays and course materials). Later on, we added the possibility to use edit distance searching in hermetic detection. By using edit distance search, it is possible to detect forms of plagiarism other than verbatim copying. The system has a Java-based user interface that enables the user to manage tasks and documents, run detection processes and browse detection results. Figure 1 illustrates the system architecture.

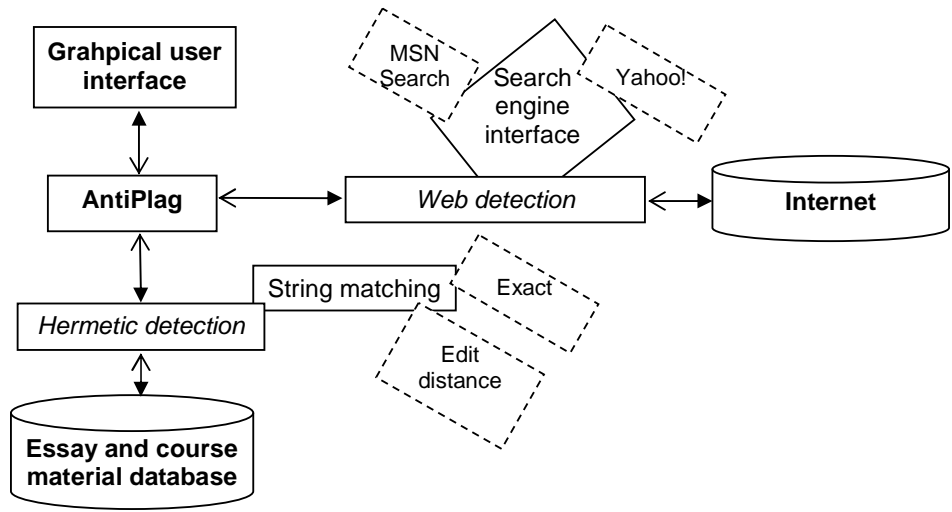

**Figure 1**: The architecture of AntiPlag. Internet detection uses "standard" web search engines MSN Live Search and Yahoo!. Hermetic detection is currently based on exact and edit distance string matching.

The current version of AntiPlag incorporates the idea of sampling-based web plagiarism detection. The system offers the user two parameters for affecting the way the sampling is performed. The *window size* (*W*) determines the number of words included in each sampling query. The *step size* (*S*) determines the frequency of samples. Figure 2 illustrates the idea.

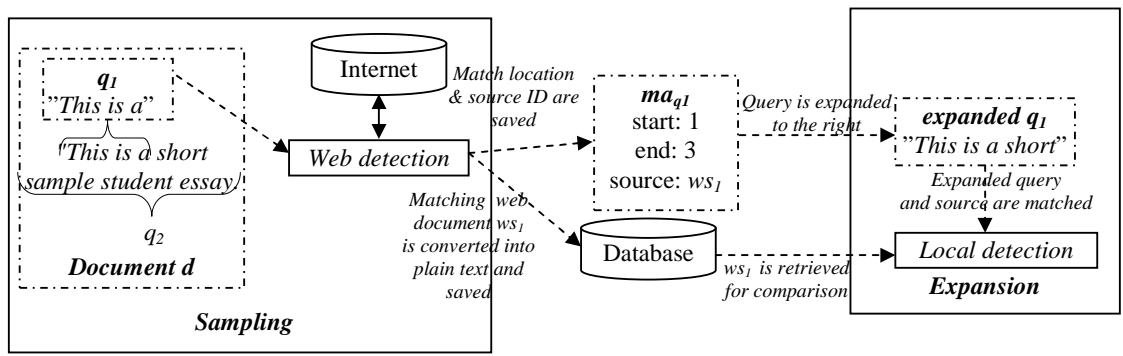

**Figure 2**: Sampling-based web plagiarism detection in AntiPlag. In this example, the two queries, $q_1$ and $q_2$, are of the window size 3 and were obtained using the S value 4.

As illustrated in Figure 2, in the *sampling phase* queries $q_n$ are generated from document *d*. Each $q_{1...n}$ is sent to the web detection component. If there is a match between a query $q_x$ and one ore more web sources $ws_i$, a *match area* ($ma_{q_x}$) is created that contains the start and end index of the query and the identifiers of the matching web sources $ws_i$. Web sources that appear in more than one match area are saved into the local database as plain text. AntiPlag is capable of converting to ASCII text MS Word, MS Excel, MS PowerPoint, PDF and HTML files.

After *d* has been sampled, *match area expansion* phase is launched. In the expansion phase, each $ma_{q_n}$ is processed to find the longest matching spans of text between *d* and each source $ws_i$. The idea is that instead of using the exact match capabilities of the Internet search engines, advanced local search techniques are utilized for finding the maximum spans. Using, for example, edit distance search, enables to find matches between text passages that contain plagiarism types other than verbatim copying. The query expansion is first attempted to both directions. If that fails, expansion continues either to the left or to the right until no expansions can be done. If overlapping areas are created during the expansion phase, they are combined.

In the final phase, those source documents that appear only in one match area that is shorter than 50 characters are filtered out from the results. The aim of the filtering is to lower the number of coincidental matches between document *d* and web sources. Finally, a list of inspected documents is displayed to the user with the information on the total percentage of the text that is suspected of being plagiarized in each document. If the percentage of plagiarized text in a document is equal or higher than 25%, AntiPlag considers it as a severe case of plagiarism and alerts the user. The user can then inspect the document using the graphical view that highlights the suspected instances of plagiarism. The user can open the suspected sources by clicking the links provided with each highlighted area. Figure 3 gives an example of the result view.

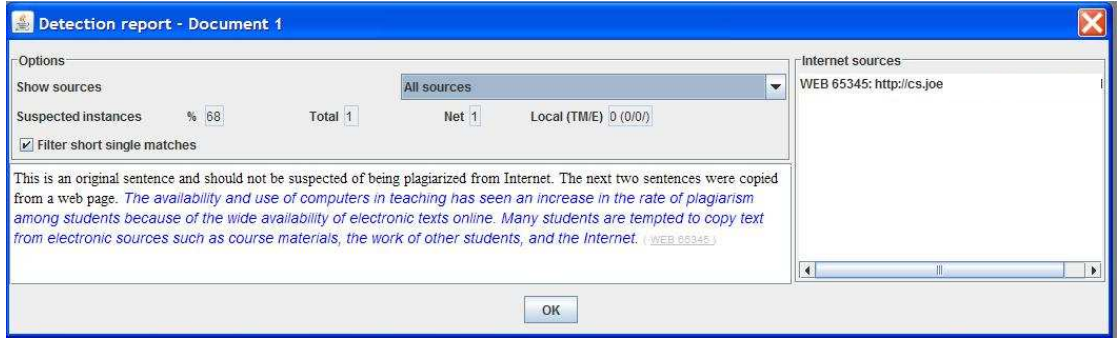

**Figure 3**: An example detection result view. The user can open the source of a suspected instance of plagiarism be clicking the link at the end of the match area.

### 4. Evaluation

We carried out an evaluation of the system on test data that consists of texts in English and has been designed for evaluating plagiarism detection systems. The data set has been used earlier for evaluating several existing plagiarism detection systems (Kakkonen & Mozgovoy 2008). Our final test set was not used during the system development phase in order to avoid the system being fitted to this particular test data.[1]

4.1 Test data and settings

The test data consists of 1,200 sentences and is divided into the following three categories: Original, Web and Mill. The sentences in the *Original* category were either deliberately written for the test set or were sourced from books that had not been published online. The *Web* category consists of sentences from a selection of Internet pages. The *Mill* sentences were obtained from *paper mill* services. Paper mills are Internet services that offer student essays for free or for payment. Each category consisted of a total of 400 sentences, 100 sentences of each test type: verbatim, edited, synonymous and paraphrased. The original, unchanged sentences from the source documents are referred to as the *verbatim sentences*. The *edited* sentences contain minor alterations such as added spaces, intentional spelling errors, deleted or added commas, and periods that were replaced by exclamation marks. In the *synonymous* sentences, one or two words were replaced with exact or close synonyms. The *paraphrased* sentences have been exposed to a wide range of sentence alterations. For instance, the kind of alterations found in the edited and synonymous sentences, and, in addition, changes in the original order of words and phrases. These test sentences formed the 48 test files, with each file containing only sentences that represented a single test type. Table 1 summarizes the properties of the test data. More details on the test set and its construction can be found in (Kakkonen & Mozgovoy 2008).

---

[1] A few bugs were identified and corrected during the extensive testing that was carried out with this final test set. Also, speed optimizations were implemented during the testing phase. These changes did not have a significant effect on the evaluation results.

**Table 1:** The content of the test data. The column "Number of files" lists the number of files in each of the test categories.

| Test category | Edit type | No. of files | No. of sentences |
|---|---|---|---|
| *Original* | Verbatim | 4 | 100 |
| | Edited | 4 | 100 |
| | Synonymous | 4 | 100 |
| | Paraphrased | 4 | 100 |
| *Web* | Verbatim | 4 | 100 |
| | Edited | 4 | 100 |
| | Synonymous | 4 | 100 |
| | Paraphrased | 4 | 100 |
| *Mill* | Verbatim | 4 | 100 |
| | Edited | 4 | 100 |
| | Synonymous | 4 | 100 |
| | Paraphrased | 4 | 100 |
| TOTAL | | 48 | 1,200 |

Separate tests were run for the files that belonged to each of the three categories (Original, Web, Mill). Each test run therefore consisted of 16 files. As none of the files in the Original test category contain instances of plagiarism, this test set was used for observing the false detection rate. All of the files in the Web and Mill sets are plagiarized and were hence used for measuring the detection accuracy.

During the development phase of the system it was found out that W value 5 offered the best compromise between the system accuracy and efficiency. In order to observe the effects of sampling frequency, the tests were repeated using the values W=6 and W=8. While using a smaller window size presumably leads to more accurate detection, it also potentially increases the number of match areas found during sampling and, hence, decreases the efficiency of the system. All the experiments were ran using Yahoo! as the web search engine and edit distance as the hermetic search method. During the development phase, 5% of the total length of the query (two or less characters for queries shorter than 40 characters) was defined as the default value for the allowed amount of edit operations in edit distance searches. All the experiments were ran using this default setting.

4.2 Results

Table 2 gives the results of the evaluation. The average accuracy over all the test categories and types was 91.7 for the window size 8 and 95.8 for the window size 6. The only category in which AntiPlag made mistakes was the Mill paraphrase category, in which the accuracy was 0% and 50%, respectively.

**Table 2:** Evaluation results. The results are reported as the percentage of files that were correctly detected either as not containing (Original) or containing alarming levels of plagiarism (>25%) (Web and Mill).

|  |  | W = 5, S = 8 | W = 5, S = 6 |
|---|---|---|---|
| Original | Verbatim | 100.0 | 100.0 |
|  | Edited | 100.0 | 100.0 |
|  | Synonymous | 100.0 | 100.0 |
|  | Paraphrase | 100.0 | 100.0 |
|  | AVERAGE | 100.0 | 100.0 |
| Web | Verbatim | 100.0 | 100.0 |
|  | Edited | 100.0 | 100.0 |
|  | Synonymous | 100.0 | 100.0 |
|  | Paraphrase | 100.0 | 100.0 |
|  | AVERAGE | 100.0 | 100.0 |
| Mill | Verbatim | 100.0 | 100.0 |
|  | Edited | 100.0 | 100.0 |
|  | Synonymous | 100.0 | 100.0 |
|  | Paraphrase | 0.0 | 50.0 |
|  | AVERAGE | 75.0 | 87.5 |
| AVERAGE |  | 91.7 | 95.8 |

4.3 Discussion

Paper (Kakkonen & Mozgovoy 2008) reported an evaluation study of four plagiarism detection systems. The same test data that we used for evaluating AntiPlag was used in these experiments. This allows us to compare our system to these four detectors. The four systems in that study were: *SafeAssignment* (Sciworth Inc 2009) is a system that can perform both local and Internet detection. *TurnitIn* (iParadigms 2009), which arguably is the most widely used of all current plagiarism detection systems, performs web and hermetic detection. While these two systems are used online, *EVE2* (version 2.5) (Canexus Inc 2009) and *Plagiarism-Finder* (PF) (version 1.3.0) (Mediaphor Software Entertainment AG 2009), in contrast, are installed on user's computer. They both support only web detection.

AntiPlag compares favorably to these four systems. The best-performing system in the experiment reported in (Kakkonen & Mozgovoy 2008) was SafeAssignment. Its average score over all the test categories and types was 87.5. It was the only system that made no mistakes on Original and Web categories. SafeAssignment scored the average of 62.5 on the Mill data. The only test category in which the system overperformed AntiPlag was the Mill paraphrase set when AntiPlag's query window size was set to 8. The second-best performer in the experiment reported in (Kakkonen & Mozgovoy 2008), TurnitIn, scored the average score of 72.9 over all the test categories. Table 3 summarizes the overall results of the system comparison.

**Table 3:** Comparison of the overall evaluation results of the five systems.

| System | Avg. accuracy |
|---|---|
| AntiPlag, W=5, S=6 | 95.8 |
| AntiPlag, W=5, S=8 | 91.7 |
| SafeAssignment | 87.5 |
| TurnitIn | 72.9 |
| EVE2 | 33.3 |
| Plagiarism-Finder | 33.3 |

**5. Conclusion**

We have introduced the AntiPlag plagiarism detection system that is aimed at web and hermetic detection from student texts. In this paper, we concentrated mostly on a particular aspect of the system, namely the sampling-based web detection method. The results of our evaluation showed that the system is more accurate than four of the existing commercial solutions. The only test category in which AntiPlag did not reach 100% accuracy was the paraphrased Mill data.

Our long-term aim is to develop AntiPlag into an advanced student plagiarism detection system capable of detecting complex forms of plagiarism. One of the most important future developments will be the improvement of the efficiency of the system. For instance, it is possible to optimize the way the match area expansion and edit distance search are implemented. An analysis of the inadequacies of existing plagiarism detection systems yielded the following four main research questions for the future work:
- How can advanced document comparison methods (such as Latent Semantic Analysis (Landauer et al 1998) be applied for the purposes of plagiarism detection?
- How can accurate stylistic "fingerprints" created from student texts and be utilized in plagiarism detection?
- How can the results obtained from an automatic reference and citation tracking system be most effectively utilized in AntiPlag? I.e. how the use of citations and references can be taken into consideration in plagiarism detection.